\documentclass[a4paper,11pt]{article}
\pdfoutput=1 

\usepackage{jheppub} 
\usepackage[T1]{fontenc}
\usepackage{booktabs}
\usepackage{xspace}
\usepackage{dcolumn}
\usepackage{hyperref}
\usepackage{caption}
\usepackage
[subrefformat=parens,position=top,skip=-15pt,margin=15pt,justification=justified,singlelinecheck=false]
{subcaption}
\usepackage{slashed}
\usepackage{comment}
\usepackage[colorinlistoftodos, shadow]{todonotes}
\usepackage{amsmath}
\usepackage{graphicx}

\makeatletter
\newcommand\myparagraph{\@startsection{paragraph}{4}{\z@}%
  {-10\p@ \@plus -6\p@ \@minus -3\p@}%
  {3\p@}%
  {\normalfont\itshape}%
}
\makeatother

\newcommand{\id}{{\rm 1\kern-.12em
 \rule{0.3pt}{1.5ex}\raisebox{0.0ex}{\rule{0.1em}{0.3pt}}}}

\allowdisplaybreaks

\def\refeq#1{\mbox{(\ref{#1})}}

\def\refse#1{\mbox{Section~\ref{#1}}}

\def\refse#1{\mbox{Section~\ref{#1}}}

\def\citere#1{\mbox{Ref.~\cite{#1}}}
\def\citeres#1{\mbox{Refs.~\cite{#1}}}

\newcommand{\cmm}[1]{\ensuremath{#1}\ifmmode\else{}\fi}
\newcommand{\nmc}[2]{\newcommand{#1}{\cmm{#2}}}
\nmc{\al}{\alpha}
\nmc{\be}{\beta}
\nmc{\de}{\delta}
\nmc{\De}{\Delta}
\nmc{\si}{\sigma}
\nmc{\eps}{\epsilon}
\newcommand{\bfphi}{\text{\boldmath{$\phi$}}}

\newcommand{\bftheta}{\text{\boldmath{$\theta$}}}
\newcommand{\bfzeta}{\text{\boldmath{$\zeta$}}}

\newcommand{\ri}{\mathrm i}
\newcommand{\ru}{\mathrm u}
\newcommand{\ie}{{i.e.}\ }

\nmc{\rd}{\mathrm{d}}

\def\beq{\begin{equation}}
\def\eeq{\end{equation}}

\newcommand{\PH}{\ensuremath{\text{H}}\xspace}
\newcommand{\PHone}{\ensuremath{{\text{H}_1}}\xspace}
\newcommand{\PHtwo}{\ensuremath{{\text{H}_2}}\xspace}
\newcommand{\PHpm}{\ensuremath{\text{H}^\pm}\xspace}
\newcommand{\PHa}{\ensuremath{A_0}\xspace}

\newcommand{\Ph}{\ensuremath{\text{h}}\xspace}

\newcommand{\Pb}{\ensuremath{\text{b}}\xspace}

\newcommand{\Pt}{\ensuremath{\text{t}}\xspace}

\newcommand{\PW}{\ensuremath{\text{W}}\xspace}
\newcommand{\PZ}{\ensuremath{\text{Z}}\xspace}

\newcommand{\MH}{\ensuremath{M_\PH}\xspace}
\newcommand{\Mh}{\ensuremath{M_\Ph}\xspace}

\newcommand{\MW}{\ensuremath{M_\PW}\xspace}

\newcommand{\MZ}{\ensuremath{M_\PZ}\xspace}

\newcommand{\GeV}{\ensuremath{\,\text{GeV}}\xspace}

\newcommand{\ren}{{\mathrm{R}}}
\newcommand{\bare}{{\mathrm{B}}}

\newcommand{\EW}{{\mathrm{EW}}}

\newcommand{\rw}{{\mathrm{w}}}
\nmc{\alem}{\alpha_{\mathrm{em}}}
\nmc{\sw}{s_{\rw}}
\nmc{\cw}{c_{\rw}}

\nmc{\g}{g_2}
\nmc{\gy}{g_1}
\nmc{\Gf}{G_\mathrm{F}}

\newcommand{\rT}{\ensuremath{\text{T}}\xspace}

\nmc{\rB}{{\rm B}}
\nmc{\rs}{{\rm s}}

\nmc{\Msb}{M_{\rm sb}}

\nmc{\ftone}{t_{s}}
\nmc{\tab}{t_{\alpha\beta}}

\nmc{\tHoneHone}{t_{H_1H_1}}
\nmc{\tHoneHtwo}{t_{H_1H_2}}
\nmc{\tHtwoHtwo}{t_{H_2H_2}}
\nmc{\tHaHa}{t_{\Ha\Ha}}
\nmc{\tHpmHpm}{t_{H^\pm H^\pm}}
\nmc{\tGzGz}{t_{G_0G_0}}
\nmc{\tGpmGpm}{t_{G^\pm G^\pm}}
\nmc{\tGzHa}{t_{G_0\Ha}}
\nmc{\tGpmHpm}{t_{G^\pm H^\pm}}

\def\draftdate{\relax}
\def\mda{\relax}
\def\mua{\relax}
\def\mla{\relax}
\def\draft{
\def\thtystars{******************************}
\def\sixtystars{\thtystars\thtystars}
\typeout{}
\typeout{\sixtystars**}
\typeout{* Draft mode!
         For final version remove \protect\draft\space in source file *}
\typeout{\sixtystars**}
\typeout{}
\def\draftdate{\today}
\def\mua{\marginpar[\boldmath\hfil$\uparrow$]%
                   {\boldmath$\uparrow$\hfil}%
                    \typeout{marginpar: $\uparrow$}\ignorespaces}
\def\mda{\marginpar[\boldmath\hfil$\downarrow$]%
                   {\boldmath$\downarrow$\hfil}%
                    \typeout{marginpar: $\downarrow$}\ignorespaces}
\def\mla{\marginpar[\boldmath\hfil$\rightarrow$]%
                   {\boldmath$\leftarrow $\hfil}%
                    \typeout{marginpar: $\leftrightarrow$}\ignorespaces}
\def\Mua{\marginpar[\boldmath\hfil$\Uparrow$]%
                   {\boldmath$\Uparrow$\hfil}%
                    \typeout{marginpar: $\uparrow$}\ignorespaces}
\def\Mda{\marginpar[\boldmath\hfil$\Downarrow$]%
                   {\boldmath$\Downarrow$\hfil}%
                    \typeout{marginpar: $\downarrow$}\ignorespaces}
\def\Mla{\marginpar[\boldmath\hfil{$\Rightarrow$}]%
                   {\boldmath{$\Leftarrow $}\hfil}%
                    \typeout{marginpar:$\leftrightarrow$}\ignorespaces}
\def\muanick{\marginpar[\boldmath\hfil$\uparrow$]%
                   {\boldmath$\textcolor{blue}\uparrow$\hfil}%
                    \typeout{marginpar:\textcolor{blue} $\uparrow$}\ignorespaces}
\def\mdanick{\marginpar[\boldmath\hfil$\downarrow$]%
                   {\boldmath$\textcolor{blue}\downarrow$\hfil}%
                    \typeout{marginpar: $\downarrow$}\ignorespaces}
\def\mlanick{\marginpar[\boldmath\hfil$\rightarrow$]%
                   {\boldmath$\textcolor{blue}\leftarrow $\hfil}%
                    \typeout{marginpar: $\leftrightarrow$}\ignorespaces}
\overfullrule 5pt
\oddsidemargin 15mm
\marginparwidth 29mm
}

\nmc{\Hhat}{\hat H}
\nmc{\hhat}{\hat h}
\nmc{\Phihat}{\hat \Phi}
\nmc{\phihat}{\hat \phi}
\nmc{\chihat}{\hat \chi}
\nmc{\etahat}{\hat \eta}
\nmc{\rhohat}{\hat \rho}
\nmc{\thetahat}{\hat \theta}
\nmc{\sihat}{\hat\sigma}
\nmc{\Phione}{\Phi_1}
\nmc{\Phitwo}{\Phi_2}
\nmc{\Zhat}{\hat Z}
\nmc{\Hahat}{\hat A_0}
\nmc{\Gzhat}{\hat G_0}
\nmc{\Xhat}{\hat X}
\nmc{\Yhat}{\hat Y}
\nmc{\Honehat}{\hat H_1}
\nmc{\Htwohat}{\hat H_2}

\nmc{\vone}{v_1}
\nmc{\vtwo}{v_2}
\nmc{\etaone}{\eta_1}
\nmc{\etatwo}{\eta_2}
\nmc{\etaplet}{\boldsymbol{\eta}}
\nmc{\chione}{\chi_1}
\nmc{\chitwo}{\chi_2}
\nmc{\chiplet}{\boldsymbol{\chi}}
\nmc{\Hone}{H_1}
\nmc{\Htwo}{H_2}
\nmc{\Hplet}{\mathbf{H}}
\nmc{\phionepm}{\phi_1^\pm}
\nmc{\phitwopm}{\phi_2^\pm}
\nmc{\Ha}{A_0}

\nmc{\tb}{t_\be}
\nmc{\ca}{c_\al}
\nmc{\catwo}{c^2_\al}
\nmc{\satwo}{s^2_\al}
\nmc{\sa}{s_\al}
\nmc{\stwoa}{s_\al}
\nmc{\cbe}{c_\be}
\nmc{\ctwobe}{c_{2\be}}
\nmc{\cbetwo}{c^2_\be}
\nmc{\sbe}{s_\be}
\nmc{\sbetwo}{s^2_\be}
\nmc{\cab}{c_{\al\be}}

\nmc{\dth}{\delta t_{H}}
\nmc{\dthhat}{\delta t_{\hat{H}}}
\nmc{\Th}{T^{H}}
\nmc{\Thhat}{T^{\Hhat}}
\nmc{\THone}{{T}^{H_1}}
\nmc{\THtwo}{{T}^{H_2}}
\nmc{\dtHone}{{\delta t}_{H_1}}
\nmc{\dtHtwo}{{\delta t}_{H_2}}
\nmc{\dtHhatone}{{\delta t}_{\hat H_1}}
\nmc{\dtHhattwo}{{\delta t}_{\hat H_2}}
\nmc{\MHone}{{M}_\PHone}
\nmc{\MHtwo}{{M}_\PHtwo}
\nmc{\MHa}{{M}_{\PHa}}
\nmc{\MHpm}{{M}_{\text{H}^\pm}}
\nmc{\MHsone}{{M}^2_\PHone}
\nmc{\MHstwo}{{M}^2_\PHtwo}
\nmc{\MHsa}{{M}^2_{\PHa}}
\nmc{\MHspm}{{M}^2_{\PHpm}}

\newcommand{\MSbar}{\ensuremath{\overline{\text{MS}}}\xspace}

\newcommand{\FJTS}{\mathrm{FJTS}}
\newcommand{\PRTS}{\mathrm{PRTS}}
\newcommand{\GIVS}{\mathrm{GIVS}}

\newcommand{\OS}{\mathrm{OS}}
\newcommand{\PhiM}{\mathbf{\Phi}}
\nmc{\PhioneM}{\mathbf{\Phi}_1}
\nmc{\PhitwoM}{\mathbf{\Phi}_2}

\nmc{\vshift}{\bar{v}}

\title{\Large Electroweak renormalization based on gauge-invariant vacuum
expectation values of non-linear Higgs representations: 1.~Standard Model}
\subheader{\today}

\author{Stefan Dittmaier$^1$,}
\affiliation{%
        $^1$Albert-Ludwigs-Universit\"at Freiburg, %
        Physikalisches Institut, %
        79104 Freiburg, %
        Germany%
}

\author{Heidi Rzehak$^2$}
\affiliation{%
	$^2$University of T\"ubingen, %
	Institute for Theoretical Physics, %
	72076 T\"ubingen, %
	Germany%
}

\emailAdd{stefan.dittmaier@physik.uni-freiburg.de,
heidi.rzehak@itp.uni-tuebingen.de}

\abstract{The renormalization of vacuum expectation value parameters, such as $v$ in the Standard Model (SM),
is an important ingredient in electroweak renormalization, where this issue is connected to the
treatment of tadpoles. 
Tadpole coun\-ter\-terms can be generated in two different ways in the Lagrangian:
in the course of parameter renormalization, or alternatively
via Higgs field redefinitions.
The former typically leads to small corrections originating from tadpoles, but 
in general suffers from 
gauge dependences
if $\MSbar$ renormalization conditions are used for mass parameters.
The latter is free from gauge dependences, but is prone to 
very large corrections in $\MSbar$ schemes, 
jeopardizing perturbative stability in predictions.
In this paper we propose a new scheme for tadpole renormalization, 
dubbed {\it Gauge-Invariant Vacuum expectation value 
Scheme (GIVS)}, which is a hybrid scheme
of the two mentioned types, with the benefits of being gauge independent and perturbatively stable.
The GIVS is based on the gauge-invariance property of Higgs fields, and the corresponding 
parameters like $v$, in non-linear representations of Higgs multiplets.
We demonstrate the perturbative stability of the GIVS in the SM
by discussing the conversion
between on-shell and $\MSbar$ renormalized masses.}

\begin{document} 

\mbox{}\hfill FR-PHENO-2022-02

\maketitle
\flushbottom

\section{Introduction}
\label{se:Introduction}

After roughly one decade of data taking, very many analyses carried out by the LHC collaborations
ATLAS and CMS in fact produce precision measurements of cross sections and model parameters
at a level where electroweak (EW)
corrections play an important role.
In the calculation of EW corrections, the procedure of renormalization is of
crucial importance, involving issues like the choice of renormalization scheme and input parameters.
Electroweak renormalization was worked out in the 1970--90s in various 
variants~\cite{Ross:1973fp,Sirlin:1980nh,Aoki:1982ed,Bohm:1986rj,Jegerlehner:1991dq,%
Denner:1991kt,Denner:1994xt}
and meanwhile is a standard procedure in modern next-to-leading order (NLO) calculations
(see, e.g., the review~\cite{Denner:2019vbn} for details and further original
references).

In some EW renormalization schemes, details of the 
parametrizations of vacuum expectation values (vevs) of
Higgs fields play a role in higher-order calculations, in others they do not.
Technically, these details concern the treatment of tadpole diagrams or subdiagrams,
which have only one external leg and can be interpreted as contributions to 
vevs of the corresponding external field.
In on-shell (OS) renormalization schemes, 
for instance, all renormalized model parameters are directly 
related to physical observables, so that in predictions for other observables 
all tadpole contributions cancel 
between loop diagrams and coun\-ter\-terms.
In renormalization schemes, however, in which not all mass parameters are tied to physical observables,
such as in schemes with $\MSbar$-renormalized mass parameters, the treatment of tadpole 
contributions has an impact on the parametrization of predicted observables in terms of
renormalized parameters and on the running behaviour of the $\MSbar$ masses.
An analogous statement also applies to models with extended Higgs sectors
in which Higgs mixing angles may be 
renormalized with $\MSbar$ conditions.

In the literature two completely different types of tadpole treatments are in use.
Both types fully absorb explicit tadpole diagrams upon introducing a tadpole coun\-ter\-term 
$\delta {\cal L}_{\delta t}=\delta t\, h$
in the Lagrangian for each Higgs field component $\phi(x)=v+h(x)$ that might acquire a non-vanishing 
vev~$v$. 
This means $\delta t$ is adjusted in such a way that the 
vev of $h$ vanishes, $\langle0|h(x)|0\rangle=0$.
The {\it explicit} tadpole terms absorbed by $\delta t$ are redistributed to coun\-ter\-terms of
other couplings as {\it implicit} tadpole contributions.
In the two variants for the tadpole treatment, however,
the tadpole coun\-ter\-terms $\delta {\cal L}_{\delta t}$ are generated in a very different way.
One possibility is to include the renormalization constant $\delta t$ in the parameter
renormalization transformation which expresses bare parameters in terms of renormalized parameters,
as, e.g., done in \citeres{Bohm:1986rj,Denner:1991kt}. We call this scheme
{\it Parameter Renormalized Tadpole Scheme (PRTS)} in the following.
The renormalized parameter~$v$ quantifying the Higgs 
vev then corresponds to the value of $\phi$ in the 
minimum of the renormalized (corrected) Higgs potential.
This scheme, however, has the unpleasant feature that $\delta t$, which is gauge dependent in general, 
enters the relations between bare parameters of the Higgs potential, and this results in a potentially
gauge-dependent parametrization of predicted observables if $\MSbar$ masses or Higgs mixing
angles are used (see discussions in \citeres{Krause:2016oke,Denner:2016etu,Denner:2018opp}).

The second possibility to introduce $\delta {\cal L}_{\delta t}$ is the {\it Fleischer--Jegerlehner
Tadpole Scheme (FJTS)} as introduced in \citere{Fleischer:1980ub},%
\footnote{This scheme is equivalent to the $\beta_t$ scheme of
\citere{Actis:2006ra}.}
where the field $h(x)$ is redefined by a transformation $h(x)\to h(x)+\Delta v$ 
with a constant $\Delta v$ in the bare
Lagrangian. This transformation is 
only a change in parametrization of the functional integral of
the generating functional of Green functions and does not change any physical prediction.
Choosing the constant $\Delta v=-\delta t/M_\Ph^2$, with $\Mh$ denoting the mass of the corresponding
Higgs boson, implies $\langle0|h(x)|0\rangle=0$ as demanded. 
The FJTS redistributes the tadpole corrections to other coun\-ter\-terms without
changing the parametrization of observables at all, i.e.\ the scheme produces the same result as if
just including all explicit tadpole diagrams wherever they appear. In other words, in this scheme
the perturbative expansion of $\phi(x)$ proceeds about the bare value of $v$, which is
related to the minimum of the bare Higgs potential, not the corrected one. 
On the upside, the FJTS does not introduce gauge dependences in the parametrization of
predictions, but by experience issues with perturbative stability potentially exist
if OS conditions for masses are not used entirely.
Comparing corrections to observables calculated within renormalization schemes with
$\MSbar$ mass parameters that differ only by using the PRTS or FJTS, the differences 
reflect the additional correction 
$\Delta v$ required in the FJTS to shift the expansion point
of $\phi$ to the renormalized value of $v$.
These additional corrections appearing in the FJTS can be very large and
jeopardize the perturbative stability of predictions.
In the SM, for instance, 
the differences between OS- and $\MSbar$-renormalized masses are unnaturally large~\cite{Jegerlehner:2012kn}.
For the top-quark mass this difference is of the order of $10\GeV$ and thus
of similar size as the QCD correction 
(see \citeres{Jegerlehner:2012kn,Kniehl:2015nwa,Kataev:2022dua} and references therein).
In models with $\MSbar$-renormalized mixing angles in extended Higgs sectors such as the
Two-Higgs Doublet Model, the FJTS is very prone to huge corrections
in extreme parameter scenarios, as discussed in 
\citeres{Krause:2016oke,Denner:2016etu,Altenkamp:2017ldc,Altenkamp:2017kxk,Denner:2018opp}.

In this paper, we propose a new tadpole scheme, 
dubbed {\it Gauge-Invariant Vacuum expectation value Scheme (GIVS)},
which aims at unifying the good 
features of the PRTS and FJTS. It is designed to expand Higgs fields about
the corrected minimum of the effective Higgs potential, so that no additional
corrections arise from correcting the expansion point. 
We, thus, expect that the GIVS 
produces only moderate corrections in predictions, very similar to the PRTS
and in contrast to the FJTS. 
This expectation is confirmed in the explicit phenomenological example discussed below.
To avoid any gauge dependences in tadpole
contributions, the construction of the GIVS makes use of non-linear
parametrizations of the Higgs 
fields~\cite{Lee:1972yfa,Grosse-Knetter:1992tbp}.
In these non-linear representations
CP-even neutral components of
Higgs multiplets---and thus also their potential constant contributions---are
gauge invariant, and the Higgs potentials are completely free of 
would-be Goldstone-boson fields, so that tadpole renormalization constants
become gauge independent. Simply using these tadpole coun\-ter\-terms
in actual calculations based on linear Higgs 
representations, would
not lead to a full compensation of explicit tadpole diagrams. 
This mismatch can be resolved upon generating a second type of tadpole
coun\-ter\-term by field shifts
as in the FJTS, so that all explicit tadpole diagrams
are cancelled by the sum of the two types of tadpole coun\-ter\-terms,
which renders the GIVS a hybrid scheme.
We formulate the GIVS for the SM and demonstrate its perturbative
stability by evaluating the differences between OS- and $\MSbar$-renormalized masses of SM
particles. In a forthcoming publication we will apply the GIVS to non-standard Higgs
sectors, e.g., with additional singlet or doublet scalar fields.

The article is organized as follows:
In \refse{se:Representations} we describe the non-linear Higgs 
representation of the SM in detail,
including the calculation of the tadpole constants in the linear and non-linear
representations. Moreover, we generally show the gauge independence of tadpoles
in the non-linear representation there.
The formulation of the GIVS as well as the 
application to the conversion of masses between OS and $\MSbar$ definitions
are presented in \refse{se:GIVscheme}.
Our conclusions are given in \refse{se:Conclusions},
and the appendix briefly describes the application of the GIVS within the
background-field method.

\section{Linear and non-linear Higgs representations of the Standard Model}
\label{se:Representations}

In this section, we introduce the linear and non-linear Higgs representations for 
the SM. All parameters and fields are considered as ``bare'' in this section,
i.e.\ the renormalization transformation for introducing renormalized quantities and renormalization
constants, including the choice of the tadpole scheme,
will be the next step after this section.
In the formulation of the SM in the linear Higgs representation 
and the definition of 
field-theoretical quantities we consistently follow the
notation and conventions of \citere{Denner:2019vbn}.
The transition to the non-linear Higgs representation uses 
\citeres{Lee:1972yfa,Grosse-Knetter:1992tbp,Dittmaier:1995cr,Dittmaier:1995ee}
as guideline.

\subsection{Kinetic Higgs Lagrangian}

Most commonly, the SM Higgs doublet is introduced as a complex two-component field $\Phi$
with charge conjugate 
$\Phi^{\mathrm{c}}=\ri\sigma_2\Phi^*$, where $\sigma_j$ ($j=1,2,3$) are denoting the
Pauli matrices in the following.
In the conventions of \citere{Denner:2019vbn}, $\Phi$ and $\Phi^{\mathrm{c}}$ are parametrized
according to
\begin{align}\label{eq:PhiSM}
\Phi = \left( \begin{array}{c}
\phi^{+} \\ \frac{1}{\sqrt2}(v +\eta +\ri\chi)
\end{array} \right),
\qquad
\Phi^{\mathrm{c}} = \left( \begin{array}{c}
\frac{1}{\sqrt2}(v +\eta -\ri\chi) \\ -\phi^{-}
\end{array} \right),
\end{align}
with the complex would-be Goldstone-boson fields $\phi^+$, $\phi^-=(\phi^+)^*$, 
the real would-be Goldstone-boson field $\chi$, the
physical Higgs field $\eta$ (called $H$ in \citere{Denner:2019vbn}), and the constant $v$
parametrizing the 
vev of the Higgs doublet.
For the transition from the linear to the non-linear Higgs representation it is convenient
to first switch to the ($2\times2$) matrix notation for the Higgs doublet 
\begin{align}\label{eq:PhilinSM}
\PhiM \equiv \left( \Phi^{\mathrm{c}},\Phi \right)
= \frac{1}{\sqrt{2}}\bigl[(v + \eta) \id + 2\ri \bfphi\bigr], \qquad
\bfphi \equiv \frac{\phi_{j} \sigma_j}{2} = \frac{\vec\phi\cdot\vec\sigma}{2},
\end{align}
with the $2\times2$ unit matrix $\id$ and $\phi_j$ $(j=1,2,3)$ are the three real
would-be Goldstone-boson degrees of freedom in a more generic notation.
Note that we use summation convention over the Goldstone index $j$,
which is sometimes replaced by a vector-like notation 
$\vec\phi=(\phi_1,\phi_2,\phi_3)^\rT$, $\vec\sigma=(\sigma_1,\sigma_2,\sigma_3)^\rT$, etc., and  
boldface characters like $\bfphi$ to indicate matrix structures.
The new field components $\phi_j$ can be identified according to 
\begin{align}
\phi^\pm = \frac{1}{\sqrt{2}}(\phi_2 \pm \ri \phi_1), \qquad \chi=-\phi_3.
\end{align}
The complex square $\Phi^\dagger\Phi$
of the Higgs doublet $\Phi$, which is the field combination
entering the Higgs potential, translates into the trace of $\PhiM^\dagger\PhiM$,
\begin{align}
\Phi^\dagger\Phi = \frac{1}{2}\mathrm{tr}\bigl[\PhiM^\dagger\PhiM\bigr]
= \phi^+\phi^- + \frac{1}{2}\bigl[ (v+\eta)^2+\chi^2 \bigr].
\end{align}

The matrix field $\mathbf\PhiM$ can be parametrized in the non-linear form 
\begin{align}\label{eq:PhinonlinSM}
  \PhiM = \frac{1}{\sqrt{2}} (v + h) U(\bfzeta), \qquad
U(\bfzeta) \equiv \exp \left(\frac{2\ri \bfzeta}{v}\right), \qquad
\bfzeta \equiv \frac{\zeta_j\sigma_j}{2},
\end{align}
in which $h$ corresponds to the physical Higgs field and 
$\vec\zeta=(\zeta_1,\zeta_2,\zeta_3)^\rT$ to real would-be Goldstone-boson components. 
Since the matrix $U(\bfzeta)$ is unitary, the square of the Higgs field does not
involve would-be Goldestone-boson fields $\zeta_j$,
\begin{align}
\mathrm{tr}\bigl[\PhiM^\dagger\PhiM\bigr] = (v+h)^2,
\end{align}
so that the Higgs potential does not involve $\zeta_j$ either.
Making use of the shorthands
\begin{align}
s_\zeta \equiv \sin\left(\frac{\zeta}{v}\right), \qquad
c_\zeta \equiv \cos\left(\frac{\zeta}{v}\right), \qquad
\zeta \equiv |\vec\zeta\,| = \bigl(\vec\zeta^{\,2}\bigr)^{1/2},
\end{align}
the relations between the component fields of the two representations are explicitly given
by~\cite{Grosse-Knetter:1992tbp}
\begin{align}\label{eq:fieldrelationSM}
 \eta ={}& c_\zeta (v + h) - v 
=  h - \frac{\zeta^2}{2v} \left(1+\frac{h}{v} \right) + {\cal O}(\zeta^4),
\nonumber\\
 \vec \phi ={}& \frac{s_\zeta}{\zeta}(v + h) \vec\zeta
= \left(1 + \frac{h}{v}\right)\vec\zeta + {\cal O}(\zeta^3), 
\end{align}
where the expansions in the second equalities neglect terms that are 
of higher order in the Goldstone fields $\zeta_j$.
Our conventions are such that $(\eta,\vec\phi\,)$ and
$(h,\vec\zeta\,)$ agree up to higher powers in the Goldstone fields.
The Higgs doublet and its  charge conjugate carry weak hypercharges
$Y_{\rw,\Phi}=+1$ and $Y_{\rw,\Phi^{\mathrm{c}}}=-1$, respectively,
so that the matrix field 
$\PhiM$ transforms under SU(2)$_{\rw}\times\mathrm{U(1)}_Y$ gauge transformations as~\cite{Grosse-Knetter:1992tbp}
\begin{align}\label{eq:Phigaugetrafo}
  \PhiM &\rightarrow S(\bftheta)\,  \PhiM\, S_Y(\theta_Y) 
\end{align}
with the transformation matrices
\begin{align}\label{eq:StrafoM}
  S(\bftheta) =\exp(\ri g_2 \bftheta), \qquad
  S_Y(\theta_Y) =\exp\left(\frac{\ri}{2} g_1\theta_Y \sigma_3\right), \qquad
\bftheta \equiv \frac{\theta_j\sigma_j}{2},
\end{align}
where $g_2$ and $g_1$ are the SU(2)$_{\rw}$ and the  U(1)$_Y$ gauge coupling constants, respectively, 
and $\vec\theta=(\theta_1,\theta_2,\theta_3)^\rT$ and $\theta_Y$ the corresponding gauge parameters.
For the field $h$ and the matrix $U(\bfzeta)$, Eq.~\refeq{eq:Phigaugetrafo} implies 
\begin{align}
h \rightarrow h, \qquad
U(\bfzeta) \rightarrow S(\bftheta)\,U(\bfzeta) \,S_Y(\theta_Y).
\label{eq:hUtrafoSM}
\end{align}
Hence, the Higgs field $h$ is invariant under gauge transformations 
while the 
fields $\zeta_j$ change under gauge transformations in a non-trivial way. 
Note also that the parameter $v$, which quantifies the Higgs 
vev, is directly associated to the gauge-invariant component $h$, 
in contrast to the linear parametrization, where the constant $v$ is attributed to the field
component $\eta$, which is not gauge invariant.

In the non-linear representation 
the Lagrangian for the kinetic terms of the Higgs and Goldstone fields
is given by
\begin{align}
{\cal L}_{\text{H,kin}} &=
\frac{1}{2}\text{tr}\left[(D_\mu \PhiM)^\dagger(D^\mu \PhiM)\right], 
\end{align}
where $D_\mu$ is the covariant derivative acting in matrix notation as
\begin{align}\label{eq:Dmu}
 D^\mu \PhiM = \partial^\mu \PhiM
-  \ri g_2  {\bf W}^\mu  \PhiM - \ri g_1 \PhiM B^\mu \frac{\sigma_3}{2},\qquad
{\bf W}^\mu \equiv \frac{W_j^\mu\sigma_j}{2},
\end{align}
with the matrix-valued SU(2)$_{\rw}$ gauge field ${\bf W}^\mu$ and
the U(1)$_Y$ gauge field $B^\mu$.
For later purposes, it is very convenient to define the following
combination of gauge fields,
\begin{align}
{\bf C}^\mu \equiv {\bf W}^\mu + \frac{g_1}{g_2} B^\mu \frac{\sigma_3}{2},
\qquad
\vec C^\mu = \left( W_1^\mu, W_2^\mu, \frac{Z^\mu}{\cw} \right)^\rT,
\end{align}
with $Z^\mu$ denoting the Z-boson field 
and $\cw=\cos\theta_\rw$ the cosine of the weak mixing angle $\theta_\rw$.
Inserting the non-linear representation \refeq{eq:PhinonlinSM} of $\PhiM$ into
${\cal L}_{\text{H,kin}}$, terms with arbitrary powers
of Goldstone fields $\zeta_j$ emerge. This complicated structure becomes
rather transparent after introducing the matrix fields
\begin{align}
{\bf W}^{(\ru)}_\mu \equiv U(\bfzeta)^\dagger \,{\bf W}_\mu \, U(\bfzeta)
+\frac{\ri}{g_2} U(\bfzeta)^\dagger \, \partial_\mu U(\bfzeta),
\qquad
{\bf C}^{(\ru)}_\mu \equiv
{\bf W}^{(\ru)}_\mu + \frac{g_1}{g_2} B_\mu \frac{\sigma_3}{2},
\end{align}
which absorb the complete $\bfzeta$ dependence of ${\cal L}_{\text{H,kin}}$,
\begin{align}
{\cal L}_{\text{H,kin}} &{}=
\frac{1}{4}\text{tr}\left[
\left(\partial_\mu h+\ri g_2 {\bf C}^{(\ru)}_\mu (v+h) \right)
\left(\partial^\mu h-\ri g_2 {\bf C}^{(\ru),\mu} (v+h)
\right)
\right]
\nonumber\\
&{}= \frac{1}{2}(\partial h)^2 + \frac{g_2^2}{8} (v+h)^2 \,
{\vec C}^{(\ru)}_\mu \cdot{\vec C}^{(\ru),\mu}.\label{eq:LSMkin}
\end{align}
The field ${\bf W}^{(\ru)}_\mu$ is identical with the gauge field ${\bf W}_\mu$
in the unitary gauge after performing the corresponding field transformation
in the full Lagrangian~\cite{Lee:1972yfa,Grosse-Knetter:1992tbp}.
The full dependence of ${\cal L}_{\text{H,kin}}$ on the Goldstone
fields can be easily derived from the components of the matrix fields 
${\bf W}^{(\ru)}_\mu$ and ${\bf C}^{(\ru)}_\mu$, which are given by
\begin{align}\label{eq:WuCu}
{\vec W}^{(\ru)}_\mu ={}& \vec W_\mu + \vec G_\mu, \qquad
{\vec C}^{(\ru)}_\mu = {\vec C}_\mu + \vec G_\mu,
\nonumber\\
\vec G_\mu ={}& 
- \frac{2s_\zeta c_\zeta}{g_2\zeta}\, \partial_\mu\vec\zeta
+ \frac{2}{g_2} \left( \frac{v s_\zeta c_\zeta}{\zeta} - 1 \right)
        \frac{(\vec\zeta\cdot\partial_\mu\vec\zeta\,)}{v\zeta}\,
	\frac{\vec\zeta}{\zeta}
- \frac{2s_{\zeta}^2}{g_2\zeta^2}\,\vec\zeta\times\partial_\mu\vec\zeta
\nonumber\\
& {} -\frac{2s_\zeta^2}{\zeta^2} \left[\zeta^2 \vec W_\mu
    -(\vec W_\mu\cdot\vec\zeta\,)\,\vec\zeta \,\right]
-\frac{2s_\zeta c_\zeta}{\zeta}\,\vec W_\mu\times\vec\zeta
\nonumber\\
={}& 
- \frac{2}{g_2 v} \, \partial_\mu\vec\zeta
- \frac{2}{g_2 v^2} \, \vec\zeta\times\partial_\mu\vec\zeta
- \frac{2}{v^2} \left[\zeta^2\,\vec W_\mu
	- \bigl(\vec W_\mu\cdot\vec\zeta\,\bigr)\,\vec\zeta\, \right]
-\frac{2}{v}\,\vec W_\mu\times\vec\zeta
\nonumber\\
& {} + {\cal O}(\zeta^3),
\end{align}
where $\times$ denotes the usual cross product of 3-dimensional vectors.
Accordingly, an expression for ${\vec C}^{(\ru)}_\mu \cdot{\vec C}^{(\ru),\mu}$ in Eq.~\eqref{eq:LSMkin} can be derived as
\begin{align}
{\vec C}^{(\ru)}_\mu \cdot{\vec C}^{(\ru),\mu}
&=  {\vec C}_\mu \cdot{\vec C}^{\mu}
- \frac{ 4 c_\zeta s_\zeta}{g_2 \zeta}\,
    (\vec C_\mu\cdot \partial^\mu \vec \zeta\,) 
+ \frac{4}{g_2} \left( \frac{ v c_\zeta s_\zeta}{\zeta}\, - 1\right)
      \frac{(\vec \zeta \cdot \vec C_\mu)(\vec \zeta \cdot \partial^\mu \vec\zeta\,)}{v\zeta^2}
\nonumber \\
&\quad {}
 - \frac{4 s_\zeta^2}{g_2 \zeta^2} (\vec C_\mu - 2 \vec W_\mu\,) \cdot (\vec \zeta \times \partial^\mu \vec \zeta\,)
\nonumber \\
&\quad
+ \frac{4g_1 s_\zeta^2}{g_2 \zeta^2}B_\mu 
      \left[ -W^\mu_3\zeta^2 +(\vec W^\mu\cdot\vec\zeta\,)\zeta_3 \right]
  - \frac{ 4 c_\zeta s_\zeta}{\zeta} \,
           \vec C_\mu\cdot ( \vec W^\mu\times\vec \zeta \,)
%
%
\nonumber \\
&\quad
 + \frac{4 s_\zeta^2}{g_2^2 \zeta^2} \,
     (\partial_\mu  \vec \zeta\,) \cdot (\partial^\mu  \vec \zeta\,)
-  \frac{4}{g_2^2}\left( \frac{v^2 s_\zeta^2}{\zeta^2} - 1 \right)
          \frac{(\vec \zeta \cdot \partial_\mu\vec \zeta\,)
             (\vec \zeta \cdot \partial^\mu \vec\zeta\,)}{v^2\zeta^2}.
\end{align}
To facilitate the derivation of Feynman rules, it is convenient to insert the expanded
version of ${\vec C}^{(\ru)}_\mu$ into ${\cal L}_{\text{H,kin}}$ up to the desired order
in $\zeta_j$ fields. Up to quadratic order in $\zeta_j$, the Higgs kinetic Lagrangian reads
\begin{align}
{\cal L}_{\text{H,kin}} &{}=
\frac{1}{2}(\partial h)^2 
+ \frac{(v+h)^2}{2 v^2}
\biggl\{
(\partial_\mu \vec \zeta\,)\cdot(\partial^\mu \vec \zeta \,)
+ \frac{g_2^2 v^2}{4}\, \vec C_\mu \cdot \vec C^\mu
\nonumber \\
& \hspace*{2em}
+g_1 g_2 B_\mu \left[ - W^\mu_3 \zeta^2
+  (\vec W^\mu \cdot \vec \zeta\,) \zeta_3 \right]
-g_2^2 v \,\vec C_\mu\cdot(\vec W^\mu \times \vec \zeta\,) 
\nonumber \\
& \hspace*{2em}
- g_2 v \,\vec C_\mu\cdot \partial^\mu \vec \zeta 
- g_2 (\vec C_\mu-2\vec W_\mu)\cdot (\vec \zeta \times \partial^\mu \vec \zeta\,)
\biggr\}
 + {\cal O}(\zeta^3).
\end{align}
The gauge-fixing Lagrangian for
the non-linear representation is given by
\begin{align}
  \mathcal L_{\text{gf}} =& - \frac{1}{2 \xi_A}\left(\partial_\mu A^\mu\right)^2  - \frac{1}{2 \xi_A}\left(\partial_\mu Z^\mu + \xi_Z \MZ \zeta_3\right)^2 
\nonumber \\
&  - \frac{1}{\xi_W}\left(\partial_\mu W^{+\,\mu} - \text{i} \xi_W \MW \zeta^+\right)\left(\partial_\mu W^{-\,\mu} + \text{i} \xi_W \MW \zeta^-\right)
  \label{eq:Lgaugefix}
\end{align}
with $\zeta^\pm = (\zeta_2 \pm \text{i} \zeta_1)/\sqrt{2}$,
the fields $W^\pm_\mu = (W_{1,\mu} \mp \text{i} W_{2,\mu})/\sqrt{2}$ corresponding to the
W$^\pm$~bosons, and $A_\mu$ being the photon field.
The parameters $\xi_a$ ($a=A,Z,W$) are the usual arbitrary gauge parameters,
and $M_{\mathrm{V}}$ ($\text{V=Z,W}$) the Z- and W-boson masses.

\subsection{Higgs potential and tadpoles}
\label{se:HpotTadSM}

The complex Higgs doublet field $\Phi$ undergoes
self-interactions as described by the Higgs potential
\begin{align}
V = -\mu_{2}^2 \big(\Phi^{\dag}\Phi\big)
+\frac{\lambda_2}{4}\big(\Phi^{\dag}\Phi\big)^2,
\label{eq:VSM}
\end{align}
with $\mu_2^2$ and $\lambda_2$ being real, positive free parameters.%
\footnote{In order to avoid a clash of notation with the reference mass scale $\mu$ of
dimensional regularization, we denote the parameters in the potential 
$\mu_2^2$ and $\lambda_2$, with the ``2'' hinting on the Higgs doublet.}
In the linear Higgs representation, the potential $V$
involves would-be Goldstone-boson fields which is not the case
in the non-linear representation, where it reads
\begin{align}
V = -\frac{\mu_{2}^2}{2} \mathrm{tr}\bigl[\PhiM^{\dag}\PhiM\bigr]
+\frac{\lambda_2}{16}\bigl(\mathrm{tr}\bigl[\PhiM^{\dag}\PhiM\bigr]\bigr)^2
= -\frac{\mu_{2}^2}{2}(v+h)^2 +\frac{\lambda_2}{16} (v+h)^4.
\end{align}
In lowest order, the vev parameter $v$ is chosen such that the field
configurations $\eta\equiv0$ and $h\equiv0$ correspond to the classical minimum
of $V$, so that no terms linear in $\eta$ or $h$ remain in~$V$.

Beyond lowest order, however, loop corrections induce non-vanishing
contributions $T^{\eta}$ and $T^{h}_{\mathrm{nl}}$ to the
one-point vertex functions $\Gamma^{\eta}$ and $\Gamma^{h}_{\mathrm{nl}}$,
known as {\it tadpole constants}.
To prevent any confusion w.r.t.\ to the two types of Higgs 
representations, we
mark vertex functions 
$\Gamma^{\dots}$ and tadpole contributions $T^{\dots}$ in the non-linear
representation by a suffix ``nl'' throughout.
For later purposes, 
we compute and compare these one-loop tadpole contributions 
$T^{\eta}$ and $T^{h}_{\mathrm{nl}}$ of the physical Higgs
fields $\eta$ and $h$ of the linear and non-linear representations, respectively.
In the linear representation $\Gamma^\eta$ is given by
\begin{align}\label{eq:Teta}
\Gamma^\eta ={} & T^\eta = 
\raisebox{-1.45ex}[1.45ex]{\hbox{\includegraphics[page=1,scale=1.2]{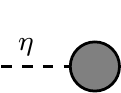}}} 
\nonumber \\[2.5ex]
= {}&\; \raisebox{-1.65ex}[1.65ex]{\hbox{\includegraphics[page=1,scale=1.2]{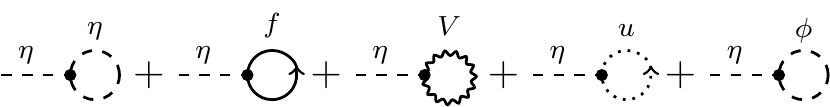}}} 
\nonumber \\
={}& \frac{1}{16 \pi^2 v}\biggl\{\frac{3}{2} \MH^2 A_0(\MH^2) - 4 \sum_f N^{\mathrm{c}}_f m_f^2 A_0(m_f^2)  
+ \MZ^2 \bigl[3 A_0(\MZ^2) - 2 \MZ^2 \bigr] 
\nonumber \\
&{} 
+ 2 \MW^2\bigl[3 A_0(\MW^2) -2 \MW^2 \bigr]
+ \frac{1}{2}\MH^2 A_0(\xi_Z \MZ^2) + \MH^2 A_0(\xi_W \MW^2)\biggr\},
\end{align}
with $\MH$ denoting the Higgs-boson mass and $m_f$ the mass of the
fermion~$f$, which has colour multiplicity $N^{\mathrm{c}}_f$.
The field label $u$ generically stands for Faddeev--Popov ghost fields.
Here we made use of the scalar 
one-point one-loop integral in $D=4-2\eps$ dimensions,
\begin{align}\label{eq:A0}
A_0(m^2) =
\displaystyle\frac{(2\pi\mu)^{4-D}}{\ri\pi^{2}}\int \rd^{D}q\, \frac{1}{q^2-m^{2}+\ri0}
= m^2\left[ \Delta + \ln\left(\frac{\mu^2}{m^2}\right) + 1 \right] + {\cal O}(\eps)
\end{align}
with the arbitrary reference mass $\mu $ and the standard UV divergence
\beq\label{eq:Delta}
\Delta = \frac{2}{4-D} + \ln 4\pi - \gamma_{\mathrm{E}},
\eeq
in which $\gamma_{\mathrm{E}}$ denotes the Euler--Mascheroni constant.
In the non-linear representation $\Gamma^h_{\mathrm{nl}}$ is given by
\begin{align}\nonumber\\[-1ex]
  \Gamma^h_{\mathrm{nl}} ={}& T^h_{\mathrm{nl}} 
= \raisebox{-1.45ex}[1.45ex]{\hbox{\includegraphics[page=1,scale=1.2]{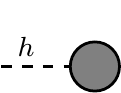}}} 
\nonumber \\[2.5ex]
={}& \raisebox{-1.6ex}[1.6ex]{\hbox{\includegraphics[page=1,scale=1.2]{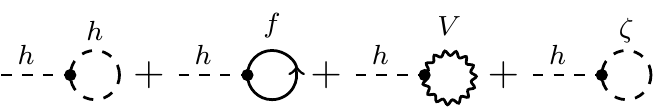}}}\nonumber \\
  ={}&\frac{1}{16 \pi^2 v}\biggl\{\frac{3}{2} \MH^2 A_0(\MH^2) - 4 \sum_f N^{\mathrm{c}}_f m_f^2 A_0(m_f^2)  
  + \MZ^2 \bigl[3 A_0(\MZ^2) - 2 \MZ^2 \bigr] 
\nonumber \\& {}
+ 2 \MW^2\bigl[3 A_0(\MW^2) -2 \MW^2 \bigr]\biggr\}.
\label{eq:Th}
\end{align}
Note that no ghost loops contribute to the tadpole $T^h_{\mathrm{nl}}$
in the non-linear representation, because the field~$h$ does not couple to 
ghost fields, since Goldstone fields $\zeta_j$ and the gauge-invariant Higgs
field~$h$ do not mix under gauge transformations. 
As already anticipated in the introduction, the gauge invariance of $h$ in the non-linear
representation has the consequence that the corresponding tadpole contribution 
$T^h_{\mathrm{nl}}$ is
gauge independent, while $T^\eta$ is not.
For later convenience, we introduce the parameter
\begin{align}
\Delta v_\xi = \frac{T^\eta-T^h_{\mathrm{nl}}}{\MH^2} 
= \frac{1}{16 \pi^2 v}\left\{
\frac{1}{2} A_0(\xi_Z \MZ^2) +  A_0(\xi_W \MW^2)\right\},
\label{eq:Deltavxi}
\end{align}
quantifying the gauge-dependent difference between the tadpole parameters
in the linear and non-linear Higgs representations.
It is interesting to note that $\Delta v_\xi$ vanishes in the Landau
gauge, where all $\xi_a=0$. This is due to the fact that the Goldstone-boson
modes are massless in Landau gauge, so that the corresponding one-loop
tadpole diagrams lead to vanishing scaleless integrals in dimensional
regularization. We do not necessarily expect that an analogous statement
holds beyond the one-loop level.

We conclude this section by considering the gauge dependences of the tadpole
functions $\Gamma^\eta$ and $\Gamma^h_{\mathrm{nl}}$ from a more general point
of view, in particular in order to get an idea about a possible generalization 
beyond the one-loop level.
The gauge-parameter dependences of any irreducible Green function
are controlled by the so-called
{\it Nielsen identities}~\cite{Kluberg-Stern:1974iel,Nielsen:1975fs,Gambino:1999ai},
expressing the invariance under {\it extended BRST variations},
which take into account variations of gauge parameters in addition to
BRS variations of the fields.
In more detail, the derivative $\partial_{\xi_a}\Gamma^{\varphi_1\varphi_2\dots}$
of the vertex function $\Gamma^{\varphi_1\varphi_2\dots}$
of some fields $\varphi_j$ w.r.t.\ a gauge parameter $\xi_a$
is expressed in terms of
local operator insertions involving the corresponding
(Grassmann-valued) BRST sources
$\gamma_{\varphi_j}$ and $\chi_a$ for the fields $\varphi_j$ and the
gauge parameter $\xi_a$, respectively.
The general formulation of the Nielsen identities within the SM,
using conventions very close to ours, can be found in
\citere{Gambino:1999ai}.
In particular, the gauge-parameter dependence of the tadpole vertex function
$\Gamma^H$ (with $H$ generically
denoting the Higgs field) is derived in Sect.~3 there, with the result
\begin{align}
\partial_{\xi_a}\Gamma^H = -\Gamma^{\chi_a\gamma_H H}(0)\Gamma^H
- \Gamma^{\chi_a\gamma_H}(0)\Gamma^{HH}(0),
\label{eq:Nielsen-tadpole}
\end{align}
where the arguments 0 on the vertex functions with more than one external leg
express the fact that external momenta are zero.
Note that this result does not depend on the Higgs representation 
(linear or non-linear), which has to be specified when calculating the 
occuring vertex functions.
The functions $\Gamma^{\chi_a\gamma_H H}$ and $\Gamma^{\chi_a\gamma_H}$ 
on the r.h.s.\ have external legs corresponding to
the sources $\chi_a$ and $\gamma_H$, so that their explicit calculation
requires extra Feynman rules involving $\chi_a$ or $\gamma_H$.
For the linear Higgs representation, these Feynman rules 
can be read off Eqs.~(A8) and (A7) of \citere{Gambino:1999ai}. 
Rewriting Eq.~\refeq{eq:Nielsen-tadpole} in terms of the field $\eta$
(to indicate the linear Higgs representation) and specializing it to the
one-loop level, we obtain
\begin{align}
\partial_{\xi_a}\Gamma^\eta = \MH^2 \Gamma^{\chi_a\gamma_\eta}(0), 
\label{eq:Nielsen-tadpole-1loop}
\end{align}
where we have used the lowest-order relations
$\Gamma^{\eta}_0=0$ and $\Gamma^{\eta\eta}_0(0)=-\MH^2$
and the fact that $\Gamma^{\chi_a\gamma_H H}$ and $\Gamma^{\chi_a\gamma_H}$
are one-loop induced.
A very simple one-loop calculation explicitly yields
\begin{align}
\partial_{\xi_Z}\Gamma^\eta ={}& \MH^2\cdot
\raisebox{-5.8ex}[5ex][8ex]{\hbox{\includegraphics[page=1,scale=1.2]{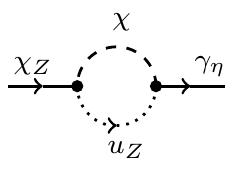}}}
= \frac{g_2 \MZ\MH^2}{64\pi^2\cw}\,B_0(0,\xi_Z\MZ^2,\xi_Z\MZ^2),
\\
\partial_{\xi_W}\Gamma^\eta ={}& \MH^2\cdot
\raisebox{-5.8ex}[5ex][5ex]{\hbox{\includegraphics[page=1,scale=1.2]{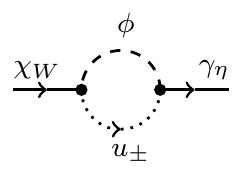}}}
=\frac{g_2 \MW\MH^2}{32\pi^2}\,B_0(0,\xi_W\MW^2,\xi_W\MW^2),
\label{eq:xi-dep-Gamma-eta}
\end{align}
with the scalar one-loop two-point function
\begin{align}\label{eq:B0}
B_0(p^2,m_0^2,m_1^2) =
\displaystyle\frac{(2\pi\mu)^{4-D}}{\ri\pi^{2}}\int \rd^{D}q\,
\frac{1}{(q^2-m_0^{2}+\ri0)[(q+p)^2-m_1^{2}+\ri0]}.
\end{align}
Making use of
\begin{align}
B_0(0,m^2,m^2) = \partial_{m^2} A_0(m^2),
\end{align}
we can see that Eq.~\refeq{eq:xi-dep-Gamma-eta} is consistent with the
$\xi$~dependence of Eq.~\refeq{eq:Teta}.
Finally, applying Eq.~\refeq{eq:Nielsen-tadpole} to the tadpole function
$\Gamma^h_{\mathrm{nl}}$ of the non-linear Higgs representation, we
immediately get the all-order result
\begin{align}
\partial_{\xi_a} \Gamma^h_{\mathrm{nl}} = 0
\label{eq:xi-dep-Gamma-h}
\end{align}
for all gauge parameters $\xi_a$,
because the BRST variation of the gauge-invariant fields~$h$ vanishes.
The proven gauge independence of the tadpole function $\Gamma^h_{\mathrm{nl}}$
is a crucial requirement for a possible 
generalization of the GIVS beyond one loop.

\section{Gauge-invariant vacuum expectation value renormalization}
\label{se:GIVscheme}

\subsection{Schemes for tadpole and vacuum expectation value renormalization}

Before formulating our new proposal for handling tadpole contributions, the GIVS, 
we first recapitulate the FJTS and PRTS variants for treating tadpoles
in the linear Higgs representation in the SM.
The following description of the FJTS and the PRTS is fully equivalent to the
one given in Sect.~3.1.6 of \citere{Denner:2019vbn}, although we have switched
to a notation for the renormalization in the Higgs sector
that is closer to the treatment of extended Higgs sectors described in
\citeres{Altenkamp:2017ldc,Altenkamp:2017kxk}, in order to prepare the
generalization of the GIVS beyond the SM.
All aspects of the renormalization procedure not spelled out explicitly below,
are exactly as described in \citere{Denner:2019vbn}.

We start out by considering the Higgs potential $V$, as defined in 
\refeq{eq:VSM}, and denote
bare parameters by subscripts ``0'' and bare fields by subscripts ``B''
in the following. 
The classical ground-state configuration $\Phi_0$ minimizes $V$,
so that
\begin{align}
\Phi_0^\dagger\Phi_0  = \frac{1}{2} \mathrm{tr}\bigl[\PhiM_0^{\dag}\PhiM_0\bigr]=
\frac{2\mu_{2,0}^{2}}{\lambda_{2,0}}.
\label{eq:Phi0}
\end{align}
We separate the ground-state configuration 
$\Phi_0=(0,v_0/\sqrt{2})^\rT$ from the bare Higgs doublet $\Phi_\bare$
by introducing a {\it bare vev parameter} $v_0$, the precise definition of which 
is specific to the chosen tadpole scheme as described below,%
\footnote{The parameter $v_0$ introduced here plays the same role as the parameter 
$\bar v$ introduced in Sect.~3.1.6 of \citere{Denner:2019vbn}. In turn,
the parameter $v_0$ defined in Eq.~(122) of \citere{Denner:2019vbn}
coincides with the parameter $v_0$ introduced here only in the FJTS, but not in
the PRTS.}
\begin{align}
\Phi_\bare  = \left( \begin{array}{c}
\phi_\bare^{+} \\ \frac{1}{\sqrt2}\bigl(v_0 +\eta_\bare +\ri\chi_\bare  \bigr)
\end{array} \right). 
\label{eq:PhiB}
\end{align}

Higher-order corrections contain tadpole diagrams, 
\ie Feynman diagrams containing subdiagrams of the
form given in Eq.~\refeq{eq:Teta}.
The vertex functions, defined via a Legendre transformation from the
connected Green functions, involve such tadpole contributions if the
splitting $v_0+\eta_\bare(x)$ of the physical Higgs field 
does not provide an expansion of the effective Higgs potential about its true
minimum (see for instance App.~C of \citere{Denner:2018opp}).  
Technically, it is desirable to organize the
perturbative bookkeeping by appropriate parameter and field definitions and
renormalization in such a way that the occurrence of tadpole
contributions is widely suppressed.
Choosing $v_0$ such that $v_0^2=4\mu_{2,0}^{2}/\lambda_{2,0}$ at least to leading order 
avoids tadpole contributions at tree level.
We will always assume this in the following.
In higher orders, the explicit (unrenormalized)
tadpole contribution $T^\eta$ of \refeq{eq:Teta}
can be cancelled upon generating a tadpole coun\-ter\-term 
$\delta t\, \eta$ in the coun\-ter\-term Lagrangian $\delta {\cal L}$.
This is achieved by a 
tadpole renormalization condition for the renormalized 
one-point function $\Gamma_{\ren}^\eta$ (in momentum space)
of the physical Higgs field,
\begin{align}
\Gamma_{\ren}^\eta = T^\eta + \delta t \overset{!}{=} 0
\quad\Rightarrow\quad
\de t = - T^\eta.
\label{eq:tadCT}
\end{align}
The tadpole coun\-ter\-term is generated by appropriately choosing $v_0$
and, if needed, by a further redefinition of the bare Higgs field $\eta_\bare$.
Inserting the field decomposition \refeq{eq:PhiB} into the bare
Lagrangian ${\cal L}$, produces a term $t_0\, \eta$ in ${\cal L}$ with
\begin{align}
t_0 = \frac{1}{4} v_0 \left(4\mu_{2,0}^2-\lambda_{2,0}v_0^2\right)
\label{eq:t0}
\end{align}
at the one-loop level,
where $t_0$ can be viewed as {\it bare tadpole constant}.
The tadpoles described below impose different conditions on $t_0$,
partially accompanied by appropriate field redefinitions of $\eta_\bare$,
in order to generate the desired tadpole coun\-ter\-term $\delta t h$ in the
coun\-ter\-term Lagrangian $\delta {\cal L}$.

\myparagraph{Fleischer--Jegerlehner tadpole scheme
(FJTS)~\cite{Fleischer:1980ub}:}

In the FJTS the bare tadpole constant is consistently set to zero,
\begin{align}
t_0 = 0, \qquad
v_0 = 2\sqrt{\frac{\mu_{2,0}^2}{\lambda_{2,0}}},
\label{eq:v0FJTS}
\end{align}
so that no tadpole coun\-ter\-term is introduced via parameter redefinitions,
and the bare Higgs-boson mass is fixed by
\begin{align}
M_{\PH,0}^2 = 2\mu_{2,0}^2.
\label{eq:MH0FJTS}
\end{align}
Instead, the tadpole coun\-ter\-term is introduced by an additional field
redefinition
\begin{align}
\eta_B &{} \;\to\; \eta_B + \Delta v^\FJTS
\label{eq:etashiftFJTS}
\end{align}
in the bare Lagrangian. This substitution leads to the term 
$-M_{\PH,0}^2\Delta v^\FJTS h$ in the bare Lagrangian,
where $M_{\PH,0}$ is the bare Higgs-boson mass.
Adjusting the constant $\Delta v^\FJTS$ according to
\begin{align}
\Delta v^\FJTS = -\frac{\de t^\FJTS}{\MH^2}= \frac{T^\eta}{\MH^2},
\label{eq:DvFJTS}
\end{align}
the explicit one-loop tadpole diagrams quantified by $T^\eta$ cancel against
the tadpole renormalization constant $\de t^\FJTS=-\MH^2\Delta v^\FJTS$,
as required in Eq.~\refeq{eq:tadCT}.
The field shift \refeq{eq:etashiftFJTS}
distributes tadpole renormalization constants to many
coun\-ter\-terms in $\de {\cal L}$. Each term of ${\cal L}$
containing a Higgs field $\eta_\bare$ produces such a coun\-ter\-term upon
replacing the $\eta$ leg in the Feynman rule by a factor $\De v^\FJTS$
(see, e.g., App.~A of \citere{Denner:2019vbn}).

Since the field shift \refeq{eq:etashiftFJTS} is a mere
reparametrization of the functional integral over the Higgs field, 
as long as all parameter renormalization constants are kept fixed,
this shift does not influence any physical observable, but only redistributes
terms in the calculation of observables.
Setting $\De v^\FJTS=0$ would be possible without
changing any prediction; the only difference in this variant is that
explicit tadpole diagrams are not cancelled by coun\-ter\-terms and have to be 
included in the calculation of corrections to observables.
This consideration, in particular, makes clear that in the FJTS
tadpole contributions correct for the fact that the effective 
Higgs potential is not expanded about the location of its 
minimum,
but about the minimum of the potential in lowest order,
which in the course of renormalization receives
further corrections.
For this reason, renormalization constants to mass parameters
receive tadpole corrections in the FJTS, which are rather large
by experience. 
In OS renormalization schemes these corrections
cancel in predictions, because these tadpole corrections systematically
cancel between self-energies and mass coun\-ter\-terms,
but in other renormalization schemes such as $\MSbar$ schemes this cancellation
is only partial, and large corrections typically remain.

On the positive side, the FJTS respects
the gauge-invariance requirement mentioned above.
To see this, recall that physical observables are always parametrized in a gauge-independent
way in terms of the original bare parameters of the theory, i.e.\ in terms of 
$\mu_{2,0}^2$ and $\lambda_{2,0}$ in the Higgs sector.
This gauge independence is neither disturbed by the gauge-independent reparametrization 
in terms of the parameters $v_0$ and $M_{\PH,0}$ from Eqs.~\refeq{eq:v0FJTS}and 
\refeq{eq:MH0FJTS}, nor by any pure field shift
such as the one provided by \refeq{eq:etashiftFJTS} even though $\De v^\FJTS$ is
gauge dependent.
Finally, the gauge independence of the parametrization of an observable in terms of
$v_0$ and $M_{\PH,0}$ carries over to the renormalized version of these parameters
if the corresponding renormalization constants do not introduce 
gauge dependences, which is for instance the case in OS and $\MSbar$ schemes in the FJTS.

\myparagraph{Parameter-renormalized tadpole scheme (PRTS)~\cite{Denner:1991kt}:}

The idea behind the PRTS is to achieve an expansion of the 
Higgs field about the true minimum of the renormalized effective Higgs potential 
(as obtained from the effective action after renormalization)
by appropriate relations among the parameters of the theory.
To this end, the bare parameter%
\footnote{The parameter $v_0$ of this paper is identical with
the parameters $\bar v=v$ introduced in the PRTS formulation of
Sect.~3.1.6.~(b) of \citere{Denner:2019vbn}, i.e.\
the meaning of the parameter~$v$ of this paper is different from $v$ 
in \citere{Denner:2019vbn}. In this paper $v=2\MW\sw/e$ is a shorthand for
a combination of measured quantities, while in 
Sect.~3.1.6.~(b) of \citere{Denner:2019vbn} $v$ is a gauge-dependent, UV-divergent 
auxiliary quantity.} 
\begin{align}
v_0=v + \de v
\label{eq:v0PRTS}
\end{align}
is renormalized in such a way that the renormalized parameter~$v$
is fixed by the renormalized parameters $\MW$ and $g_2=e/\sw$,
which are directly related to measured values,
\begin{align}
v = \frac{2\MW}{g_2} = \frac{2\MW\sw}{e},
\end{align}
where $\sw=\sin\theta_\rw$ is the sinus of the weak mixing angle $\theta_\rw$
and $e$ the electric unit charge.
The corresponding renormalization constant
\begin{align}
\frac{\de v}{v} = -\de Z_e + \frac{\de\sw}{\sw} +\frac{\de\MW^2}{2\MW^2},
\end{align}
is, thus, directly fixed by the renormalization conditions on
$e$, $\MW$, and $\sw^2={1-\MW^2/\MZ^2}$.
In order to guarantee the compensation of all tadpole contributions after 
renormalization, the bare tadpole constant $t_0$ given in Eq.~\refeq{eq:t0}
is split into a renormalized value $t^\PRTS$ and a corresponding
renormalization constant $\de t^\PRTS$,
\begin{align}
t_0 = t^\PRTS + \de t^\PRTS,
\end{align}
and demanding $t^\PRTS=0$.
On the l.h.s.\ of Eq.~\refeq{eq:t0}, this simply replaces $t_0$ by $\de t^\PRTS$,
on the r.h.s.\ the bare parameter $v_0$ is inserted according to Eq.~\refeq{eq:v0PRTS};
this leads to
\begin{align}
\delta t^\PRTS= v_0\left(\mu_{2,0}^2-\frac{1}{4}\lambda_{2,0} v_0^2\right)
= {v}\left(\mu_{2,0}^2-\frac{1}{4}\lambda_{2,0} v^2
-\frac{1}{2}\lambda_{2,0} v\de v \right),
\label{eq:dtPRTS}
\end{align}
where the second equality holds in one-loop approximation.
Since the renormalized parameter $v$, which is directly fixed by measurements, and 
the original bare parameters $\mu_{2,0}^2$ and $\lambda_{2,0}$ are gauge independent,
the gauge dependence of $\delta t^\PRTS$ goes over to $\de v$,
where it shows up as gauge dependence in the mass renormalization constant $\de\MW^2$.

In the renormalization procedure,
the two bare parameters $\mu_{2,0}^2$ and $\lambda_{2,0}$ of the Higgs sector
are tied to two renormalized parameters, for which we take $v$ as specified above and
the Higgs-boson mass $\MH$. The link to $\MH$ is provided by
the squared bare Higgs mass $M_{\PH,0}^2 =\MH^2+\de\MH^2$, where $\MH$ is fixed by experiment and the
renormalization constant $\de\MH^2$ by a renormalization condition.
The bare Higgs-boson mass $M_{\PH,0}$ is related to the bare parameters 
according to
\begin{align}
M_{\PH,0}^2 = -\mu_{2,0}^2 +\frac{3}{4}\lambda_{2,0} v_0^2
= -\mu_{2,0}^2 +\frac{3}{4}\lambda_{2,0} v^2 
+\frac{3}{2}\lambda_{2,0}  v \de v,
\label{eq:MH0PRTS}
\end{align}
where again Eq.~\refeq{eq:v0PRTS} was used in the last equality.
From Eqs.~\refeq{eq:dtPRTS} and \refeq{eq:MH0PRTS}, we see that the PRTS tadpole renormalization
constant $\delta t^\PRTS$ can also be introduced by the 
replacements~\cite{Denner:2016etu,Denner:2019vbn}
\begin{align}\label{eq:dtgenPRTS}
\lambda_{2,0}\;\to\;\lambda_{2,0}+2\frac{\de t^\PRTS}{v^3}, \qquad
\mu_{2,0}^2\;\to\;\mu_{2,0}^2+\frac{3}{2}\frac{\de t^\PRTS}{v}
\end{align}
in the bare Lagrangian with $t_0=0$.
As a result of the described procedure, 
some vertex coun\-ter\-terms receive
contributions from $\delta t^\PRTS$; the corresponding coun\-ter\-term Feynman rules
can, e.g., be found in App.~A of \citere{Denner:2019vbn}.

As mentioned before, these gauge dependences of the PRTS fully drop out in predictions
based on OS-renormalized parameters. 
If $\MSbar$-renormalized mass parameters are used as input,
the gauge dependence of $\delta t^\PRTS$ enters the parametrization of
observables in the step where $\mu_{2,0}^2$ and $\lambda_{2,0}$ are traded for
$v_0$ and $M_{\PH,0}$ via Eqs.~\refeq{eq:dtPRTS} and \refeq{eq:MH0PRTS}.
However,
these gauge dependences do not invalidate the applicability of the PRTS.
In spite of the gauge dependences,
consistent predictions can either be produced upon fixing a gauge
once and for all, or by translating measured input parameters
between different gauge choices.
By experience, the PRTS has the practical advantage over the FJTS
that contributions to mass renormalization constants are much smaller,
which, in particular, implies that conversions of renormalized 
mass parameters between OS
and $\MSbar$ renormalization schemes are typically
much smaller in the PRTS as compared to the FJTS
(see also \refse{se:MSbarmasses}).

\myparagraph{Gauge-Invariant Vacuum expectation value Scheme (GIVS):}

The aim in the new proposal of this paper is to
unify the benefits of the FJTS and the PRTS:
the gauge-invariance property of the former and the perturbative stability
of the latter.
To avoid potentially large corrections induced by tadpole loops
as inherent in the FJTS, we tie the vev of the Higgs field
to the ``true'' minimum of the effective 
Higgs potential, i.e.\ to the
Higgs potential expressed in terms of renormalized parameters,
as done in the PRTS.
The gauge dependences in the PRTS result from the fact that the 
location of the minimum
of the renormalized effective
Higgs potential, quantified by the parameter $v$,
is translated into a condition $v_0=v+\de v$ for
the non-gauge-invariant component $v_0+\eta_\bare(x)$ of the Higgs doublet~$\Phi$
\refeq{eq:PhiSM}
in the linear Higgs representation.

This problem is avoided by switching to the non-linear Higgs representation
\refeq{eq:PhinonlinSM}
where the condition $v_0=v+\de v$ applies to the gauge-invariant
component $v_0+h_\bare(x)$, a fact 
that gives the GIVS its name.
In detail, we fix the tadpole coun\-ter\-term by
\begin{align}
\delta t^\PRTS_{\mathrm{nl}} = -T^h_{\mathrm{nl}},
\end{align}
where the tadpole contribution $T^h_{\mathrm{nl}}$ results from the 
one-point function of the $h$~field in the 
non-linear Higgs representation, $\Gamma^h_{\mathrm{nl}}=T^h_{\mathrm{nl}}$.
Generating now tadpole coun\-ter\-terms from the bare Lagrangian according to
Eq.~\refeq{eq:dtgenPRTS} with $\delta t^\PRTS_{\mathrm{nl}}$ instead of 
$\de t^\PRTS$, this procedure is just the application of the PRTS in
the non-linear representation. 
Note, however, that $\delta t^\PRTS_{\mathrm{nl}}$ 
is a gauge-independent
constant, so that the PRTS in the non-linear Higgs representation
does not suffer from gauge dependences.
This procedure already fully defines the GIVS in the non-linear representation,
but almost all explicit calculations of EW corrections are carried out
in the linear Higgs representation.

The GIVS is defined in the linear Higgs representation in such a way that
the effect of tadpole renormalization is exactly the same as in the non-linear 
representation. This means that we set
\begin{align}
\de t^\GIVS_1 \equiv \delta t^\PRTS_{\mathrm{nl}} = -T^h_{\mathrm{nl}},
\end{align}
which is the (gauge-independent) part of the tadpole renormalization that goes into 
relation \refeq{eq:dtPRTS}
between bare parameters.
The tadpole coun\-ter\-terms proportional to $\de t^\GIVS_1$ are exactly the ones
as generated in the PRTS according to Eq.~\refeq{eq:dtgenPRTS}
with $\de t^\PRTS$ replaced by $\de t^\GIVS_1$.
Since, however, $T^h_{\mathrm{nl}}\ne T^\eta$, these tadpole coun\-ter\-terms are not sufficient
to cancel all explicit tadpole diagrams, which go with $T^\eta$ in the 
linear representation.
We achieve the complete cancellation of explicit tadpole diagrams upon generating
additional tadpole coun\-ter\-terms 
as in the FJTS by a field shift $\eta_B\to\eta_B+\Delta v^\GIVS$
in the bare Lagrangian
with $\de t^\GIVS_2=-\MH^2\Delta v^\GIVS$ and demand
\begin{align}
\de t^\GIVS ={}& \de t^\GIVS_1 + \de t^\GIVS_2 \overset{!}{=} -T^\eta.
\nonumber\\
\Rightarrow\;
\de t^\GIVS_2 ={}& -\MH^2\Delta v^\GIVS = T^h_{\mathrm{nl}}-T^\eta = -\MH^2 \Delta v_\xi,
\quad \mbox{i.e.\ \; $\Delta v^\GIVS =\Delta v_\xi$,}
\label{eq:SMdtGIVS}
\end{align}
with $\Delta v_\xi$ representing the gauge-dependent quantity defined in
\refeq{eq:Deltavxi}.
The constant $\de t^\GIVS_2$ is gauge dependent, but does not have any effect on
physical observables, analogously to its role in the FJTS.

To summarize, the GIVS is a hybrid version of the PRTS and the FJTS
with two types of tadpole coun\-ter\-terms:
the ones connected to 
$\de t^\GIVS_1=v_0(\mu_{2,0}^2-\lambda_{2,0} v_0^2/4)$ 
as $\de t^{\mathrm{PRTS}}$ in the PRTS 
and the ones connected to $\De v^\GIVS$ in the same way as $\De v^\FJTS$ in the FJTS.
The GIVS tadpole coun\-ter\-terms are generated from the bare
Lagrangian with 
$t_0=0$ by the substitutions
\begin{align}
\lambda_{2,0}&{} \;\to\;\lambda_{2,0}+\frac{2}{v^3}\de t^\GIVS_1,
\qquad
\mu_{2,0}^2\;\to\;\mu_{2,0}^2+\frac{3}{2v}\de t^\GIVS_1,
\nonumber
\\
\eta_B &{} \;\to\; \eta_B - \de t^\GIVS_2/\MH^2, 
\end{align}
which combines the substitutions \refeq{eq:etashiftFJTS} and \refeq{eq:dtgenPRTS}
of the FJTS and PRTS, respectively.
Alternatively,
with the tadpole coun\-ter\-terms of the FJTS and the PRTS already
generated, the generation of the one-loop GIVS tadpole coun\-ter\-terms is easily
accomplished by the substitutions
\begin{align}
\de t^{\PRTS} \;\to\; \de t^\GIVS_1, \qquad
\de t^{\FJTS} \;\to\; \de t^\GIVS_2.
\end{align}
These substitutions can, e.g., be directly applied to
the SM Feynman rules given in App.~A of \citere{Denner:2019vbn}.
If both $\de t^{\PRTS}$ and $\de t^{\FJTS}$ contribute
to a coun\-ter\-term vertex, in which case simply $\de t$ is written in those 
Feynman rules, the full GIVS tadpole constant $\de t^\GIVS$ has to be taken,
\begin{align}
\de t \;\to\; \de t^\GIVS = \de t^\GIVS_1 + \de t^\GIVS_2 = -T^\eta.
\end{align}
This is, in particular, the case for the coun\-ter\-term in the
Higgs 
one-point function $\Gamma_\ren^\eta$, which receives the coun\-ter\-term
$\de t=-T^\eta$ so that $\Gamma_\ren^\eta=0$ as demanded.

\subsection{Relation between on-shell and $\MSbar$ renormalized masses in the SM}
\label{se:MSbarmasses}

In order to compare the different tadpole renormalization schemes, 
we consider the relation between $\MSbar$ and OS renormalized masses,  
$\overline{M}$ and $M^{\OS}$, respectively. 
The link between $\overline{M}$ and $M^{\OS}$ is provided by the bare mass
parameter $M_0$, which is split into a renormalized mass and a corresponding
mass renormalization constant $\delta \overline{M}$ or $\delta M^{\OS}$
in the two schemes,
\begin{align}
M_0 = M^{\OS} + \delta M^{\OS} = \overline{M} + \delta \overline{M}.
\end{align}
Taking into account that the $\MSbar$ renormalization constant $\delta \overline{M}$
only consists of the UV-divergent contributions proportional to 
the standard UV divergence $\Delta$ defined in Eq.~\eqref{eq:Delta},
the mass difference $\Delta M^{\MSbar{-}\OS}$ is given by
\begin{align}
\Delta M^{\MSbar{-}\OS} = \overline{M} - M^{\OS}
= \delta M^{\OS} - \delta \overline{M}
= \delta M^{\OS}\big|_\text{finite},
\end{align}
where the suffix ``finite'' means that $\Delta$ is set to zero.
The mass difference, thus, can be calculated from the OS mass renormalization
constant $\delta M^{\OS}$ upon setting the UV-divergent constant $\Delta$ to zero
and specifying a value for the scale $\mu$ which now plays the role of 
a renormalization scale.
For expressing the OS constants $\delta M^{\OS}$ in terms of self-energy
functions $\Sigma(p^2)$ at on-shell points $p^2=(M^{\OS})^2$, we follow
the notation and conventions of \citere{Denner:2019vbn}
where self-energies $\Sigma(p^2)$ do not only include the contribution
$\Sigma_{1\text{PI}}(p^2)$ from one-particle irreducible (1PI) diagrams,
but also all explicit tadpole loops and tadpole counterterms
(see Eq.~(141) in \citere{Denner:2019vbn}).
Omitting the superscript ``OS'' for on-shell masses throughout,
we obtain at the one-loop level
\begin{align}
  \label{eq:convmf}
  \Delta m^{\MSbar{-}\OS}_{f} ={}&\frac{1}{2}\biggl[
 \text{Re}\Sigma^{ff,\text{r}}_{1\text{PI}}(m_f^2) 
+\text{Re}\Sigma^{ff,\text{l}}_{1\text{PI}}(m_f^2) 
+m_f \left[\text{Re} \Sigma^{ff,\text{L}}_{1\text{PI}}(m_f^2) + \text{Re} \Sigma^{ff,\text{R}}_{1\text{PI}}(m_f^2)\right] 
\nonumber\\
& {}
-  2m_f \frac{\Delta v}{v}\biggr]_{\text{finite}},
\\
\label{eq:convMV}
  \Delta M^{\MSbar{-}\OS}_{V} ={}& 
\left[\frac{1}{2}\frac{\text{Re} \Sigma^{VV}_{\text{T}, 1\text{PI}}(M_V^2)}{M_V} 
-  M_V \frac{\Delta v}{v}\right]_{\text{finite}} \qquad  V= \PZ, \PW,
\\
\label{eq:convMH}
  \Delta M^{\MSbar{-}\OS}_{\PH} ={}& 
\left[\frac{1}{2}\frac{\text{Re} \Sigma^{\eta\eta}_{1\text{PI}}(\MH^2)}{\MH} 
-  \frac{3}{2}\MH \frac{\Delta v}{v}\right]_{\text{finite}}.
\end{align}
Explicit expression for self-energy functions can, e.g., be found in
\citere{Denner:1991kt}.%
\footnote{In \citere{Denner:1991kt} the functions $\Sigma_{1\text{PI}}$ are simply 
called $\Sigma$, the scalar parts of the fermion self-energies are combined to
$\Sigma^{ff,\text{S}}=\Sigma^{ff,\text{r}}+\Sigma^{ff,\text{l}}$,
and the Higgs self-energy $\Sigma^{\eta\eta}_{1\text{PI}}$ is denoted $\Sigma^{\PH}$.}
The sum of all tadpole contributions (explicit loop diagrams and renormalization constants)
is contained in the $\Delta v$ term, which is chosen according to the applied tadpole scheme,
\begin{align}
\Delta v^{\text{FJTS}} = \frac{T^\eta}{\MH^2}, \qquad  
\Delta v^{\text{PRTS}} = 0, \qquad 
\Delta v^{\text{GIVS}} =  \frac{T^\eta - T^h_{\text{nl}}}{\MH^2} = \Delta v_\xi.
\end{align}

In Tab.~\ref{tab:Mdiff}, we list the numerical values for 
$\Delta M^{\MSbar{-}\OS}_{\EW}$ according to Eqs.~\eqref{eq:convmf}--\eqref{eq:convMH}
for the heaviest particles in the SM,
where the subscript ``EW'' indicates that we only include EW (one-loop) corrections.
\begin{table}
\centerline{
\setlength{\arraycolsep}{.5em}
$\begin{array}{|c|c|c|c|c|}
\hline
& M^\OS[\mathrm{GeV}]
& \multicolumn{3}{c|}{\vphantom{\rule{0em}{1.2em}}\Delta M^{\MSbar{-}\OS}_{\EW}[\mathrm{GeV}]}
\\[.2em]
& & \FJTS & \PRTS & \GIVS
\\
\hline
\mbox{W boson} & 80.379  & -2.22 & 0.82 & 0.74\\
\mbox{Z boson} & 91.1876 & -0.77  & 1.25 & 1.14\\
\mbox{Higgs boson} & 125.1 & 6.34 & 3.16 & 2.80 \\
\hline
\mbox{top quark} & 172.4 & 10.75 & 0.99 & 0.54\\
\mbox{bottom quark} & 4.93 & -1.79 & 0.10 & 0.13\\
\hline
\mbox{$\tau$ lepton} & 1.77686 & -0.93 & -0.028 & -0.015\\
\hline
\end{array}$
}
\caption{On-shell masses $M^\OS$ of the heaviest SM particles and differences
\mbox{$\Delta M^{\MSbar{-}\OS}_{\EW}$} between the \MSbar mass
$\overline{M}(\mu=M^\OS)$ and $M^\OS$
induced by NLO EW corrections using the FJTS, PRTS, or GIVS.}
\label{tab:Mdiff}
\end{table}
The masses entering in  $\Delta M^{\MSbar{-}\OS}_{\EW}$ are chosen according to the OS mass 
values given in Tab.~\ref{tab:Mdiff}, and the inverse of the fine-structure constant 
is chosen as $\alpha_{\mathrm{em}}^{-1} = 137.0359997$.
All other masses, i.e.\ the fermion masses of the first 
and second generations, are set to zero. The conversion is calculated at the renormalization 
scale of the corresponding OS mass, $\mu = M^\OS$
and for the gauge-dependent PRTS the 
't~Hooft--Feynman gauge ($\xi_a=1$) is chosen,
as mostly done in practice.
For the FJTS, we compared to the numerical value of the top-quark mass shift given in 
\citere{Jegerlehner:2012kn} and find agreement. The values obtained in the PRTS and 
the GIVS are of comparable size while in general the FJTS leads to larger differences 
between the OS and the \MSbar masses. An exception is the conversion of the Z-boson mass,
for which all three tadpole schemes produce mass shifts of the moderate size that is naively
expected from EW corrections.
As emphasized in the literature~\cite{Jegerlehner:2012kn, Kniehl:2015nwa, Kataev:2022dua}
for the top quark before,
the FJTS shift $\Delta m^{\MSbar{-}\OS}_{\Pt,\EW}=10.75\GeV$ in the conversion of fermion masses 
is much larger than the typical size of EW corrections 
of the percent level. 
For the lighter fermions $\Pb$ and $\tau$, the relative corrections
$\Delta M^{\MSbar{-}\OS}_{\EW}/M^\OS$ are even larger than for the top quark in the FJTS,
reaching up to $\sim50\%$, while the shifts in the PRTS and GIVS remain 
all moderate.
The large corrections in the FJTS are mostly due to the top-quark loop
in $\Delta v^{\text{FJTS}} = T^\eta/\MH^2$.
Despite these large corrections, the FJTS often is
favoured in the literature in this
context, since it leads to a gauge-independent result in contrast to the PRTS. 
Note, however, that these large EW one-loop corrections entail an enhancement of the
theoretical uncertainties due to missing higher-order corrections.
The GIVS, on the other hand, provides gauge-independent mass shifts that are
moderate and, thus, leads to smaller EW theory uncertainties, when those
uncertainties are estimated by the propagation of the known corrections to higher
order as typically done.%
\footnote{There is a large cancellation in the
mass shift $\Delta m^{\MSbar{-}\OS}_{\Pt,\EW}$ between the
one-loop QCD and EW corrections in the FJTS scheme, as pointed out in
\citere{Jegerlehner:2012kn}. Since this cancellation is, however, accidental,
it does not lead to a reduction of theoretical uncertainties from missing higher orders.}

\section{Conclusions}
\label{se:Conclusions}

Extensive discussions in the literature have shown that the
two mostly used prescriptions for tadpole contributions in
EW renormalization lead to unsatisfactory results in predictions
based on $\MSbar$ renormalization conditions. 
The tadpole prescription (called PRTS in this paper)
in which relations between parameters
are exploited to generate tadpole coun\-ter\-terms show decent 
perturbative stability, but suffer from gauge dependences;
on the other hand, generating tadpole coun\-ter\-terms from Higgs
field redefinitions (called FJTS) avoid gauge dependences, but
potentially suffers from perturbative instabilities.
The difference between the two tadpole schemes can be interpreted
as different choices of vevs for the Higgs field at higher orders,
i.e.\ the separation of the physical Higgs field into a constant contribution
and field excitation is different in the two schemes.
The PRTS expands about the ``true'' (corrected) vev, while the FJTS 
leads to potentially large corrections in the renormalization of mass parameters
originating from the perturbative shift in the Higgs vev.
In the SM, this issue
concerns $\MSbar$-renormalized mass parameters,
in models with extended Higgs sector this concerns $\MSbar$-renormalized
Higgs mixing angles in addition.

Motivated by this unsatisfactory situation, we have proposed a hybrid
scheme of the PRTS and FJTS variants unifying the strengths and avoiding the
weaknesses of the PRTS and FJTS schemes by generating the
gauge-dependent part of the tadpole coun\-ter\-term al la FJTS, where it
does not enter predictions for observables, and the potentially large
gauge-independent part a la PRTS, where it is 
absorbed into
parameter relations which in turn protects observables from large
corrections. 
The new scheme is called 
{\it Gauge-Invariant Vacuum expectation value Scheme (GIVS)},
because it exploits the fact that parameters $v_i$ determining 
Higgs vevs like the famous parameter $v$ in the SM,
appear as parts of truly gauge-invariant field components of Higgs multiplets $\Phi_i$
if these multiplets are represented in an appropriate non-linear fashion.
These non-linear Higgs representations factorize the would-be Goldstone-boson
parts from the remaining Higgs field components in such a way that
gauge-invariant  combinations of the fields $\Phi_i$, such as $\Phi^\dagger_i\Phi_j$,
do not involve Goldstone fields. 
Thus, Goldstone fields do not appear in the Higgs potential at all.
The condition that determines the $v_i$ by minimizing the 
effective Higgs potential
does not involve Goldstone fields, resulting in gauge-independent
tadpole corrections that can be absorbed into parameter relations
as in the PRTS. 
The GIVS, thus, expands Higgs fields about the ``true'' minimum of the 
effective Higgs
potential, like the PRTS, but in a representation in which the
vevs acting as expansion points are gauge invariant.
The hybrid character of the GIVS comes into play by fixing tadpole
renormalization constants $\delta t_i$ in the non-linear representation of the Higgs
sector and making use of 
these $\delta t_i$ in the linear representation
where these $\delta t_i$ are supplemented by FJTS-like contributions to
fully cancel all explicit tadpole diagrams.
We stress that actual loop calculations in the GIVS can be entirely carried out 
in the linear Higgs representations like for the PRTS and FJTS, once the
simple tadpole constants are known, i.e.\ calculations in the GIVS are not more
complicated than usual.

We have described the GIVS for the SM in such detail that further applications
of this scheme at the one-loop level should be simple.
Owing to the gauge-invariance property of the Higgs field and its
one-point function in the non-linear representation, which follows from a Nielsen
identity, we expect that the GIVS can be generalized to higher loop levels without
major obstacles.
To illustrate the perturbative stability of one-loop results 
based on 
$\MSbar$ renormalization with the GIVS tadpole treatment, we have
discussed the mass parameter conversion between OS and 
$\MSbar$-renormalized masses in the
SM. As expected, the GIVS leads to small shifts between 
$\MSbar$- and OS-renormalized
masses, in contrast to the FJTS.
     In a forthcoming publication, we will apply the GIVS to a scalar singlet extension of
the SM and to the Two-Higgs-Doublet Model and investigate the perturbative stability of
$\MSbar$ renormalization of the Higgs mixing angles. We expect that the GIVS outperforms the
FJTS in view of stability, very similar to the PRTS, but without the downside of the PRTS
of leading to gauge dependences.

\subsection*{Acknowledgements}

\begin{sloppypar}
  We thank Ansgar Denner for helpful discussions and for comments on the manuscript.
  H.R.'s research is funded by  the
Deutsche Forschungsgemeinschaft (DFG, German Research
Foundation)---project no.\ 442089526; 442089660.
\end{sloppypar}


\appendix
\section*{Appendix}

\section{The GIVS in the background-field method}

In the quantization of gauge fields via the background-field method (BFM) 
(see, e.g., \citeres{Weinberg:1996kr,Bohm:2001yx,Schwartz:2013pla}),
each field $\Psi$ 
of the conventional formalism 
is split into a {\it quantum field} $\Psi$ and
a {\it background field} $\hat\Psi$.
The quantum field represents the integration variable in the functional integral
employed for quantization
and leads to lines inside loops of Feynman diagrams in perturbative calculations;
the background field acts as (classical)
source in the effective action and leads to external lines and tree-like propagators
in Feynman diagrams.
One of the great benefits of the BFM is the gauge invariance of its effective
action $\hat\Gamma[\hat\Psi]$, which is achieved by the independent gauge fixings of the quantum
and background parts of the gauge fields.
This property leads to QED-like Ward identities for the 1PI
Green functions instead of the much
more complicated Slavnov--Taylor identities
in the conventional quantization formalism. These Ward identities in turn
imply simplifications in the renormalization procedure, as for instance
discussed in \citeres{Denner:1994xt,Denner:2019vbn} for the EW part of the SM.
In this appendix, we briefly review the formulation of the non-linear
Higgs representation in the BFM and show that the implementation of the GIVS in the BFM
is fully straightforward, i.e.\
the benefits of BFM renormalization and the GIVS can be combined easily. 

The application of the BFM 
to the SM in
the non-linear Higgs representation is described in 
\citeres{Dittmaier:1995cr,Dittmaier:1995ee}.
The splitting of fields into
quantum and background parts proceeds according to the usual linear splitting
\begin{align}
\Psi \;\to\; \tilde\Psi = \Psi+\hat\Psi
\end{align}
for all fields but the would-be Goldstone-boson fields $\zeta_j$, which are
contained in the unitary matrix $U=U(\bfzeta)$. For these fields, the splitting
proceeds multiplicatively for the matrix $U$ according to 
\begin{align}
U \;\to\; \tilde U = \hat U U, \qquad
\tilde U= U(\tilde\bfzeta),\qquad \hat U= U(\hat\bfzeta),
\end{align}
so that $\tilde\bfzeta = \hat\bfzeta+\bfzeta+{\cal O}(\zeta^2)$,
where ${\cal O}(\zeta^2)$ stands for any field monomials at least bilinear
in any combination of quantum or background would-be Goldstone-boson fields.
Exploiting the unitarity of $\tilde U$, 
the kinetic Higgs Lagrangian becomes
\begin{align}
{\cal L}_{\text{H,kin}} &{}=
\frac{1}{4}\left(v+\tilde{h}\right)^2
\mathrm{tr}\left[\left(\tilde{D}_{\mu}\tilde{U}\right)^\dagger\left(\tilde{D}^{\mu}\tilde{U}\right)\right] 
+\frac{1}{2}\left(\partial_{\mu}\tilde{h}\right)\left(\partial^{\mu}\tilde{h}\right),
\end{align}
and the Higgs potential is again independent of the would-be Goldstone-boson fields,
\begin{align}
V = -\frac{\mu_{2}^2}{2} \left(v+\tilde{h}\right)^2
+\frac{\lambda_{2}}{16} \left(v+\tilde{h}\right)^4.
\end{align}
In order to calculate 1PI Green functions, at least the gauge of the quantum
gauge fields has to be fixed. The background gauge-invariance of the 
effective action requires a special form of the gauge fixing term.
Following \citeres{Dittmaier:1995cr,Dittmaier:1995ee}, we take
\begin{align}
{\cal L}_{\mathrm{fix}}=
-\frac{1}{\xi_Q}\mathrm{tr}\left[\left(\hat{D}_{W}^{\mu}{\bf W}_{\mu}
+\frac{1}{2}\xi_Q g_{2}v\hat{U}\bfzeta\hat{U}^\dagger\right)^2\right]
-\frac{1}{2\xi_Q}\left(\partial^{\mu}B_{\mu}+\frac{1}{2}\xi_Q g_{1}v\zeta_{3}\right)^2,
\end{align}
with the covariant derivative in the adjoint representation defined as
\begin{align}
\hat{D}_{W}^{\mu}{\bf X}=\partial^\mu {\bf X}-\ri g_{2}\bigl[\hat{\bf W}^\mu,{\bf X}\bigr],
\end{align}
where ${\bf X}$ stands for any matrix-valued field transforming in the adjoint representation.
To be in line with the BFM formulation in the linear Higgs 
representation~\cite{Denner:1994xt,Denner:2019vbn},
we take a common 
gauge parameter $\xi_Q$ for both the SU(2)$_\rw$ and U(1)$_Y$
gauge fields, although it would be possible 
to introduce different gauge 
parameters for the two group factors.
The derivation of the corresponding Faddeev--Popov Lagrangian 
proceeds as usual, and the result involves neither $h$ nor 
$\hat h$ in the non-linear Higgs representation.

In the BFM, each conventional Feynman rule splits into different versions with 
different numbers of quantum and background fields; for the linear Higgs 
representation the Feynman rules are explicitly given in \citeres{Denner:1994xt,Denner:2019vbn}.
For the calculation
of 1PI Green functions (vertex functions), only Feynman rules
with exactly two quantum fields are needed, corresponding to the fact that exactly
two loop lines are attached to each vertex.
For the formulation of the GIVS, we only need the Higgs one-point function
$\Gamma^{\hat h}_{\mathrm{nl}}=T^{\hat h}_{\mathrm{nl}}$ 
in the non-linear Higgs representation, the calculation of which
only requires all $\hat h\Psi^\dagger\Psi$ terms for the quantum fields $\Psi$ of the Lagrangian
at one loop.
The intermediate steps of this calculation are straightforward and simple, so that we only quote
the result that 
the tadpole constants in the BFM completely agree
with the corresponding tadpoles of the conventional formalism after
setting all gauge parameters $\xi_a$ to $\xi_Q$,
although the break-up into bosonic diagramatic contributions of gauge bosons, would-be Goldstone
bosons, and ghost fields is somewhat different,
\begin{align}
\Gamma^{\hat h} = T^{\hat h} = T^{h}\big|_{\xi_a=\xi_Q}, \qquad 
\Gamma^{\hat h}_{\mathrm{nl}} =T^{\hat h}_{\mathrm{nl}} = T^{h}_{\mathrm{nl}}\big|_{\xi_a=\xi_Q}.
\label{eq:BFMtad}
\end{align}

The implementation of the GIVS in the BFM renormalization procedure works
exactly as described in \refse{se:GIVscheme} for the conventional formalism.
The tadpole renormalization constant $\delta t^\GIVS$ consists of the same
two parts $\de t^\GIVS_1$ and $\de t^\GIVS_2$ as defined in Eq.~\refeq{eq:SMdtGIVS}.
Nominally the tadpole constants $T^{h}_{\mathrm{(nl)}}$
have to be replaced by
$T^{\hat h}_{\mathrm{(nl)}}$, but according to Eq.~\refeq{eq:BFMtad} those
quantities do not change in the transition to the BFM for $\xi_a=\xi_Q$.
The generation of the tadpole coun\-ter\-terms follows the same strategy as 
in the conventional formalism as well. 
Making use of the BFM Feynman rules given in App.~A of
\citere{Denner:2019vbn}, the GIVS tadpole coun\-ter\-terms are obtained
by the substitutions
$\de t^{\PRTS} \to \de t^\GIVS_1$,
$\de t^{\FJTS} \to \de t^\GIVS_2$, and
$\de t \to \de t^\GIVS_1+\de t^\GIVS_2$.

\bibliographystyle{JHEPmod}
\bibliography{vevren-sm} 

\end{document}